%% file: gc_excess_lathuile2017_malyshev.tex
\documentclass{cimento}

\usepackage{xspace} 
\usepackage{amsmath}
\usepackage{amsfonts}
\usepackage{amsxtra}
\usepackage{amssymb}

\usepackage{lineno}

\def\refapj@jnl#1{{\it #1}}%
\newcommand\aj{\refapj@jnl{AJ}}
\newcommand\araa{\refapj@jnl{ARA\&A}}
\newcommand\apj{\refapj@jnl{ApJ}}
\newcommand\apjl{\refapj@jnl{ApJL}}     
\newcommand\apjs{\refapj@jnl{ApJS}}
\newcommand\ao{\refapj@jnl{ApOpt}}
\newcommand\apss{\refapj@jnl{Ap\&SS}}
\newcommand\aap{\refapj@jnl{A\&A}}
\newcommand\aapr{\refapj@jnl{A\&A~Rv}}
\newcommand\aaps{\refapj@jnl{A\&AS}}
\newcommand\azh{\refapj@jnl{AZh}}
\newcommand\baas{\refapj@jnl{BAAS}}
\newcommand\icarus{\refapj@jnl{Icarus}}
\newcommand\jrasc{\refapj@jnl{JRASC}}
\newcommand\memras{\refapj@jnl{MmRAS}}
\newcommand\mnras{\refapj@jnl{MNRAS}}
\newcommand\pra{\refapj@jnl{PhRvA}}
\newcommand\prb{\refapj@jnl{PhRvB}}
\newcommand\prc{\refapj@jnl{PhRvC}}
\newcommand\prd{\refapj@jnl{PhRvD}}
\newcommand\pre{\refapj@jnl{PhRvE}}
\newcommand\prl{\refapj@jnl{PhRvL}}
\newcommand\pasp{\refapj@jnl{PASP}}
\newcommand\pasj{\refapj@jnl{PASJ}}
\newcommand\qjras{\refapj@jnl{QJRAS}}
\newcommand\skytel{\refapj@jnl{S\&T}}
\newcommand\solphys{\refapj@jnl{SoPh}}
\newcommand\sovast{\refapj@jnl{Soviet~Ast.}}
\newcommand\ssr{\refapj@jnl{SSRv}}
\newcommand\zap{\refapj@jnl{ZA}}
\newcommand\nat{\refapj@jnl{Nature}}
\newcommand\iaucirc{\refapj@jnl{IAUC}}
\newcommand\aplett{\refapj@jnl{Astrophys.~Lett.}}
\newcommand\apspr{\refapj@jnl{Astrophys.~Space~Phys.~Res.}}
\newcommand\bain{\refapj@jnl{BAN}}
\newcommand\fcp{\refapj@jnl{FCPh}}
\newcommand\gca{\refapj@jnl{GeoCoA}}
\newcommand\grl{\refapj@jnl{Geophys.~Res.~Lett.}}
\newcommand\jcp{\refapj@jnl{JChPh}}
\newcommand\jgr{\refapj@jnl{J.~Geophys.~Res.}}
\newcommand\jqsrt{\refapj@jnl{JQSRT}}
\newcommand\memsai{\refapj@jnl{MmSAI}}
\newcommand\nphysa{\refapj@jnl{NuPhA}}
\newcommand\physrep{\refapj@jnl{PhR}}
\newcommand\physscr{\refapj@jnl{PhyS}}
\newcommand\planss{\refapj@jnl{Planet.~Space~Sci.}}
\newcommand\procspie{\refapj@jnl{Proc.~SPIE}}
\newcommand\actaa{\refapj@jnl{AcA}}
\newcommand\caa{\refapj@jnl{ChA\&A}}
\newcommand\cjaa{\refapj@jnl{ChJA\&A}}
\newcommand\jcap{\refapj@jnl{JCAP}}
\newcommand\na{\refapj@jnl{NewA}}
\newcommand\nar{\refapj@jnl{NewAR}}
\newcommand\pasa{\refapj@jnl{PASA}}
\newcommand\rmxaa{\refapj@jnl{RMxAA}}
\newcommand\maps{\refapj@jnl{M\&PS}}
\newcommand\aas{\refapj@jnl{AAS Meeting Abstracts}}
\newcommand\dps{\refapj@jnl{AAS/DPS Meeting Abstracts}}

\newcommand{\Fermi}{\textit{Fermi}\xspace}

\newcommand{\onepic}{0.37}
\newcommand{\twopic}{0.3}

\newcommand{\be}{\begin{equation}}
\newcommand{\ee}{\end{equation}}

\usepackage{graphicx}  

\title{Galactic center gamma-ray excess and the \Fermi bubbles}
\author{D.~Malyshev\from{ins:a}\from{ins:b}}
\instlist{ \inst{ins:a}  Erlangen Centre for Astroparticle Physics, Erwin-Rommel-Str. 1, D-91058 Erlangen, Germany
\inst{ins:b} on behalf of the \Fermi LAT collaboration}

\begin{document}

\maketitle

\input{gc_excess_lathuile2017_malyshev_body}

\bibliography{gce_papers_lathuile}  

\end{document}

%% file: gc_excess_lathuile2017_malyshev_body.tex
\begin{abstract}
Galactic center (GC) is expected to be the brightest source of possible dark matter (DM) annihilation signal.
Excess gamma-ray emission has been detected by several groups.
Both DM and more conventional astrophysical explanations of the excess have been proposed.
In this report, we discuss possible effects of modeling the \Fermi bubbles at low latitudes on the GC excess.
We consider a template of the \Fermi bubbles at low latitudes derived by assuming that the spectrum between 
1 GeV and 10 GeV at low latitudes is the same as at high latitudes.
We argue that the presence of the \Fermi bubbles near the GC may have a significant influence 
on the spectrum of the GC excess, especially at energies above 10 GeV.
\end{abstract}

\section{Introduction}

Galactic center (GC) is expected to be the brightest source of possible annihilation of dark matter (DM) particles \cite{2009Sci...325..970K}.
Hints of the excess in the GC were claimed
soon after the \Fermi LAT data became available \cite{Goodenough:2009gk, 2009arXiv0912.3828V}.
It was also argued that the excess has an extension larger than the \Fermi LAT point spread function \cite{Hooper:2011ti}.
Further studies by several groups have confirmed that the excess is indeed extended
with a spectrum peaking around a few GeV
\cite{2012PhRvD..86h3511A, 2013PDU.....2..118H, Gordon:2013vta, Daylan:2014rsa}.

Apart from DM annihilation, possible interpretations of the excess include additional sources of cosmic rays (CR) near the GC
\cite{2015JCAP...12..005C, 2015JCAP...12..056G, 2015arXiv151004698C},
an unresolved population of millisecond pulsars (MSPs)
\cite{2013A&A...554A..62G, 2014JHEAp...3....1Y, 2015JCAP...02..023P, 
2016PhRvL.116e1102B, 2016PhRvL.116e1103L, 2015ApJ...812...15B},
\Fermi bubbles near the Galactic plane (GP) \cite{2016A&A...589A.117Y, 2016arXiv161106644M}.
The interpretation of the GC excess depends crucially on uncertainties
related to modeling of the Galactic foreground emission and resolved point sources
\cite{Calore:2014xka, 2016ApJ...819...44A, 2017arXiv170403910T}.

One of the largest uncertainties in the GC excess is the behavior
of the \Fermi bubbles near the GP \cite{2017arXiv170403910T}.
Although the \Fermi bubbles are relatively easy to model at high latitudes above and below 
the GP \cite{2010ApJ...724.1044S, 2014ApJ...793...64A},
the study the \Fermi bubbles near the GP suffers from the same uncertainties in the Galactic foreground
modeling as the GC excess itself.
The problem is further complicated by the absence of clear counterpart of the \Fermi bubbles in
other frequencies.
Numerical modeling of the bubbles 
\cite{2012MNRAS.424..666Z, 2012ApJ...756..182G, 2012ApJ...761..185Y, 2014ApJ...790..109M}
and observations of lobes in other galaxies show that the bubbles
can be either completely expelled from the GC by the pressure of the gas,
or they can have an hourglass shape centered at the GC,
or they can have an extended base in the GC.
Although the gamma-ray spectrum of the \Fermi bubbles may change as a function of the latitude
due to energy losses, energy dependent propagation effects, or re-acceleration inside the volume of the bubbles,
observations suggest that the spectrum of the bubbles is uniform at high latitudes
\cite{2013PDU.....2..118H, 2014ApJ...793...64A}.

In this report, we consider a model of the \Fermi bubbles at low latitudes derived with the assumption that 
their spectrum is the same at low latitudes as at high latitudes
and discuss the effect of including the low latitude bubbles template on the GC excess spectrum.
The discussion is based on the results reported in \cite{2017arXiv170403910T}.

\section{GC excess with high latitudes bubbles template}
\label{sec:sample}

In order to study the effect of the \Fermi bubbles on the GC excess, we first review a derivation of the excess
in a model that includes a template for the bubbles at latitudes $|b| > 10^\circ$ \cite{2014ApJ...793...64A}.
The analysis is based on 6.5 years of {\Fermi}-LAT data between August 4, 2008  and January 31, 2015, Pass~8 UltraCleanVeto class events
with a zenith angle cut $\theta < 90^{\circ}$.
We take the data between 100 MeV and 1 TeV in 27 logarithmic energy bins.
The maps are constructed using HEALPix \cite{2005ApJ...622..759G} with a pixel size of $\approx 0^\circ\!\!.46$
at low latitudes and $\approx 0^\circ\!\!.92$ at high latitudes \cite{2017arXiv170403910T}.

\begin{figure}[htbp]
\begin{center}
\includegraphics[scale=\twopic]{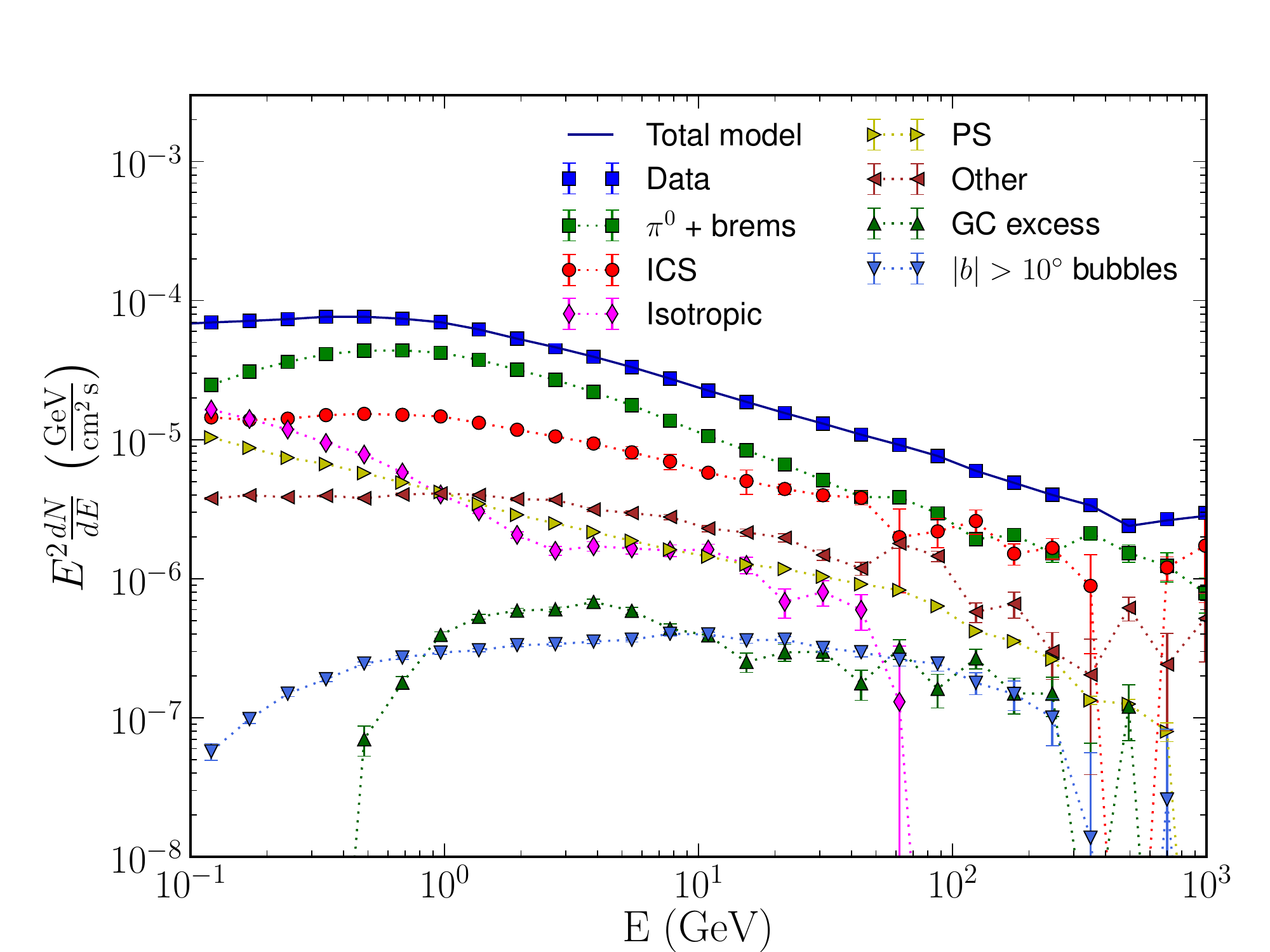}
\includegraphics[scale=\twopic]{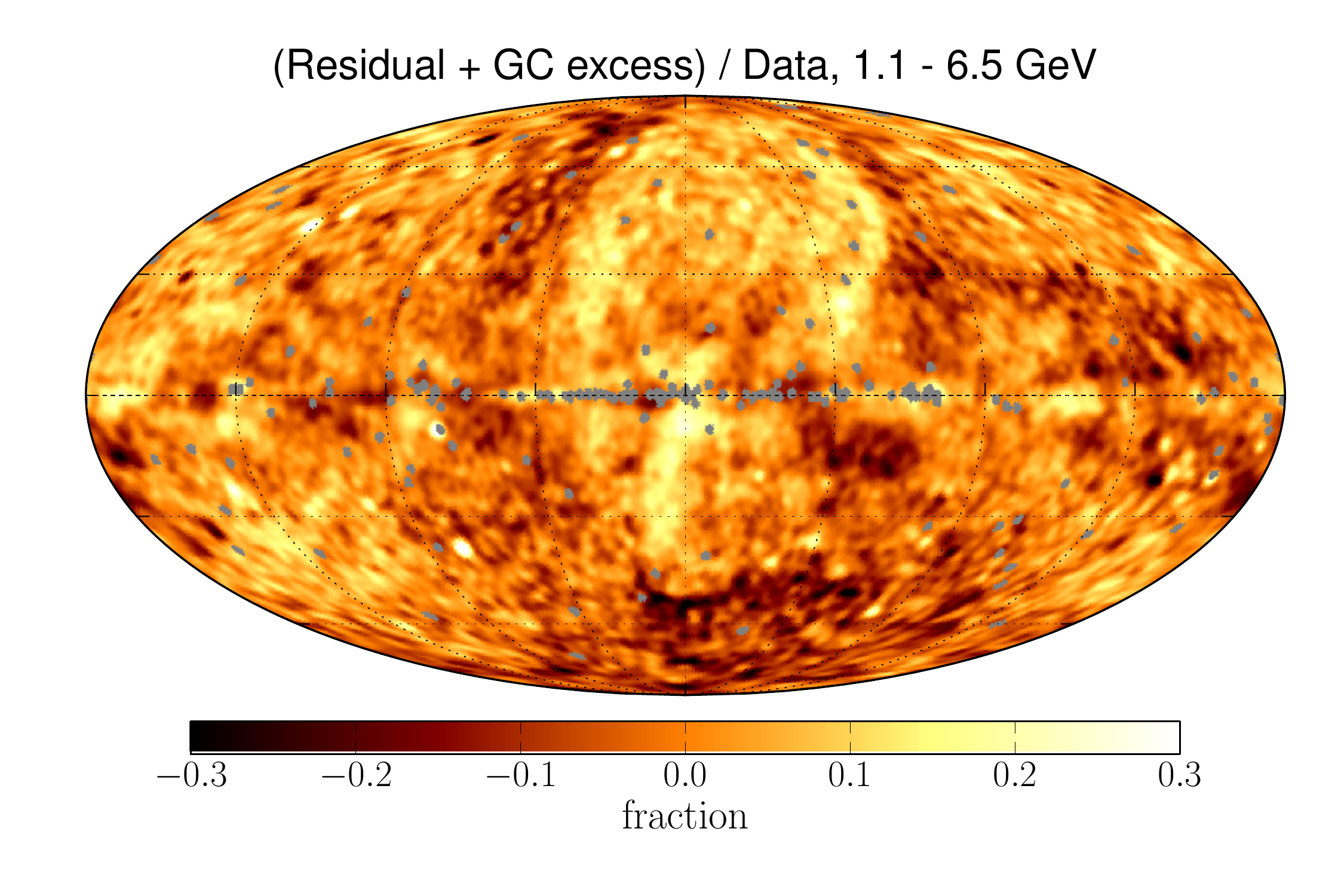}
\noindent
\caption{\small 
Left: spectrum of components of gamma-ray emission in a model with \Fermi bubbles determined at $|b| > 10^\circ$.
``Other" components include Loop I, extended sources, Sun, and Moon templates.
See Section \ref{sec:sample} for more details on the different components of emission.
Right: fractional residual in the model with spectra shown on the left plot summed over energy bins between 1.1 GeV
and 6.5 GeV plus the GC excess modeled by the gNFW DM annihilation template. 
}
\label{fig:baseline_spectra}
\end{center}
\end{figure}

The model is derived by fitting templates corresponding to different emission components to the data in each energy bin.
The templates consist of $\pi^0$ and bremsstrahlung emission components separated in five Galactocentric rings
derived with GALPROP code 
\cite{Strong:2004de, Ptuskin:2005ax, Porter:2008ve, Vladimirov:2010aq}.
The inverse Compton templates are also derived with the GALPROP code. They are separated in three components
related to the three interstellar radiation fields (cosmic microwave background, starlight, and infrared components).
The other components are Loop I, flat \Fermi bubbles template at high latitudes, Sun and Moon templates, point sources template 
(derived with fluxes from the 3FGL catalog). 
The Large Magellanic cloud, the Cygnus region, and the other extended sources in the 3FGL catalog
are treated as independent components. The cores of 200 brightest PS from the 3FGL catalog are masked within $1^\circ$.
The GC excess is modeled by DM annihilation in generalized NFW profile \cite{1996MNRAS.278..488Z} with index $\gamma = 1.25$, 
$\rho(r) \propto \frac{1}{r^\gamma (1 + r)^{3 - \gamma}}$.
The spectra of the different components obtained by fitting the templates to the data in each energy bin
and the residual with the GC excess emission added back
are shown in Figure \ref{fig:baseline_spectra}.
Although the excess flux integrated over the whole sky is relatively small, $\lesssim 1\%$,
the intensity of excess emission near the GC is about $15\%$ of the total gamma-ray intensity
in that region.

\section{\Fermi bubbles template at low latitudes}

\begin{figure}[htbp]
\begin{center}
\includegraphics[scale=\onepic]{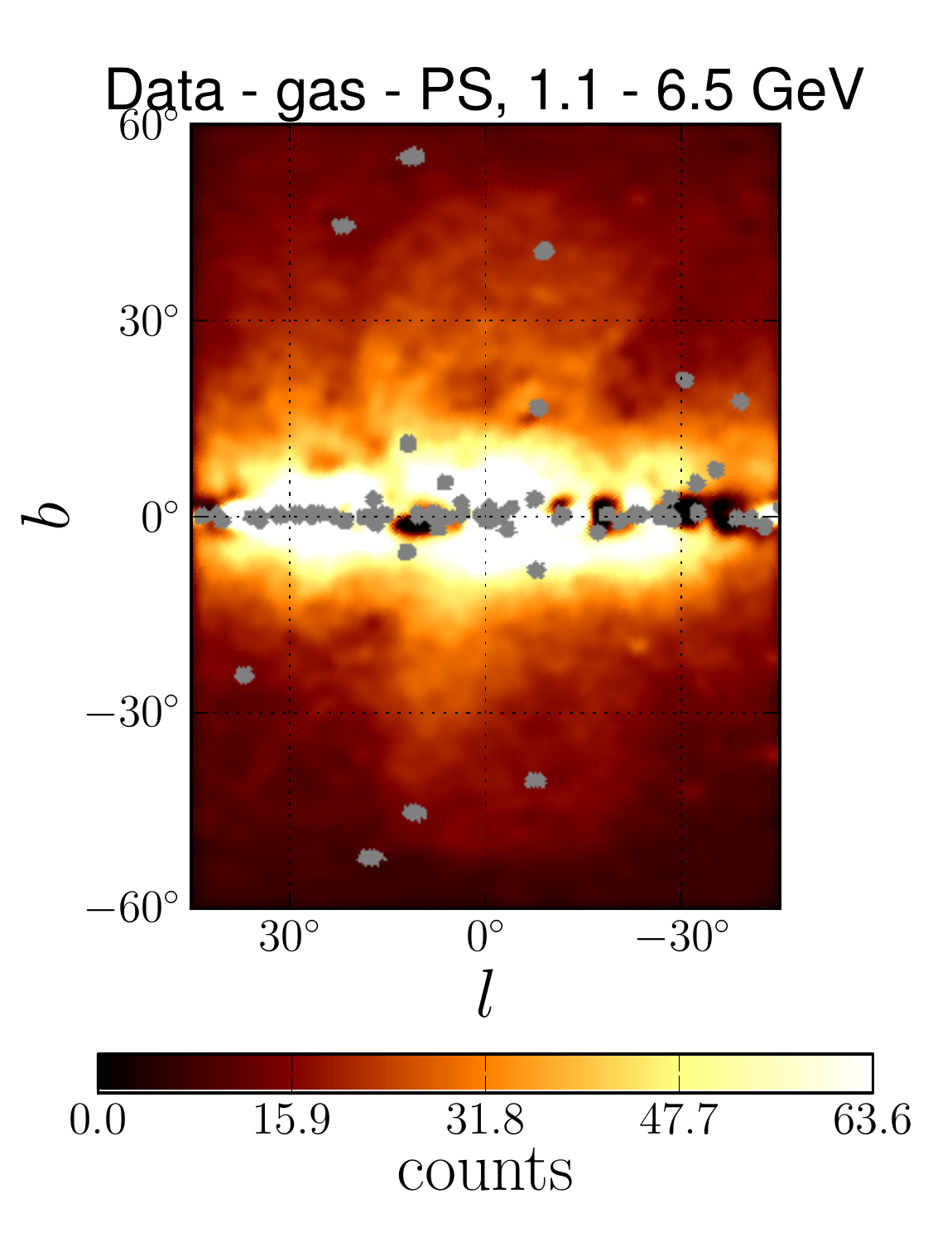}
\includegraphics[scale=\onepic]{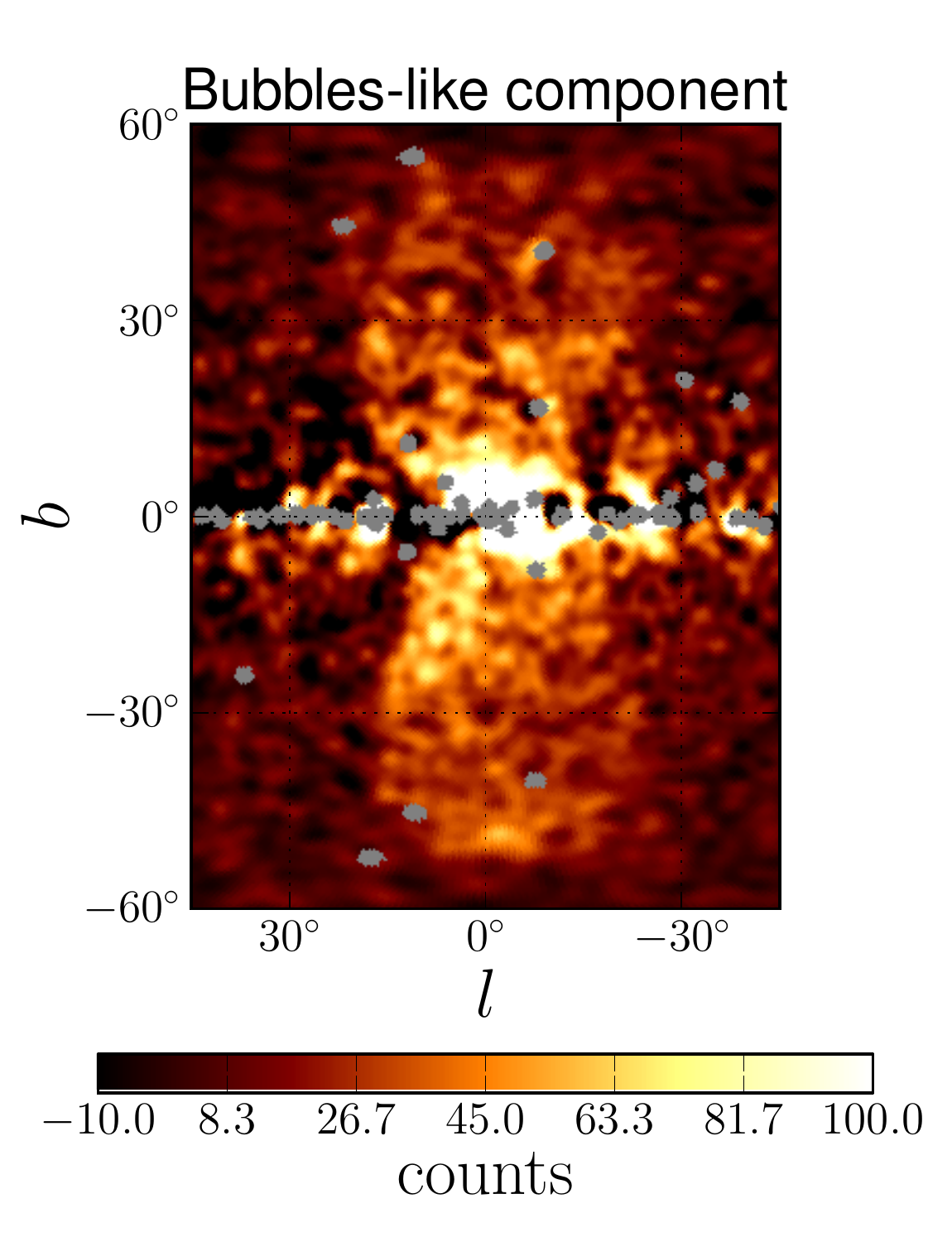}
\includegraphics[scale=\onepic]{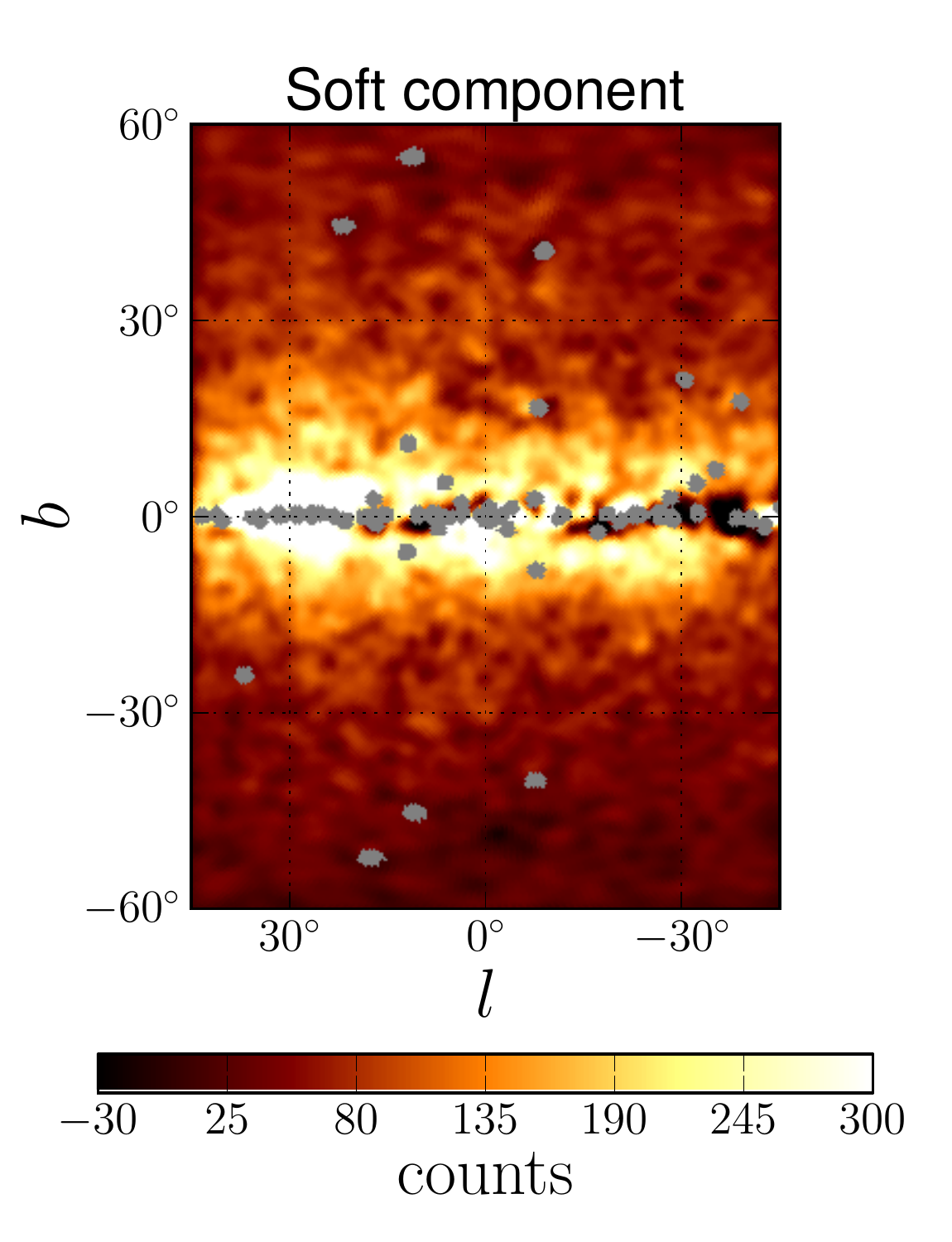}
\noindent
\caption{\small 
Left panel: residual obtained by subtracting gas-correlated components of emission 
($\pi^0$ and bremsstrahlung) and PS from the data.
These residuals are decomposed into two spectral components:
$\propto E^{-1.9}$ spectrum (middle panel) and $\propto E^{-2.4}$ spectrum (right panel).
}
\label{fig:SCA_components}
\end{center}
\end{figure}

Since there are no clear counterparts of the \Fermi bubbles in other frequencies that can be
used to derive a template of the gamma-ray emission, 
one needs to make some assumption to construct a model of the bubbles at low latitudes.
The assumption that we will make is that the spectrum of the bubbles at low latitudes is the same
as at high latitudes in the energy range between 1 GeV and 10 GeV.

\begin{figure}[htbp]
\begin{center}
\includegraphics[scale=\onepic]{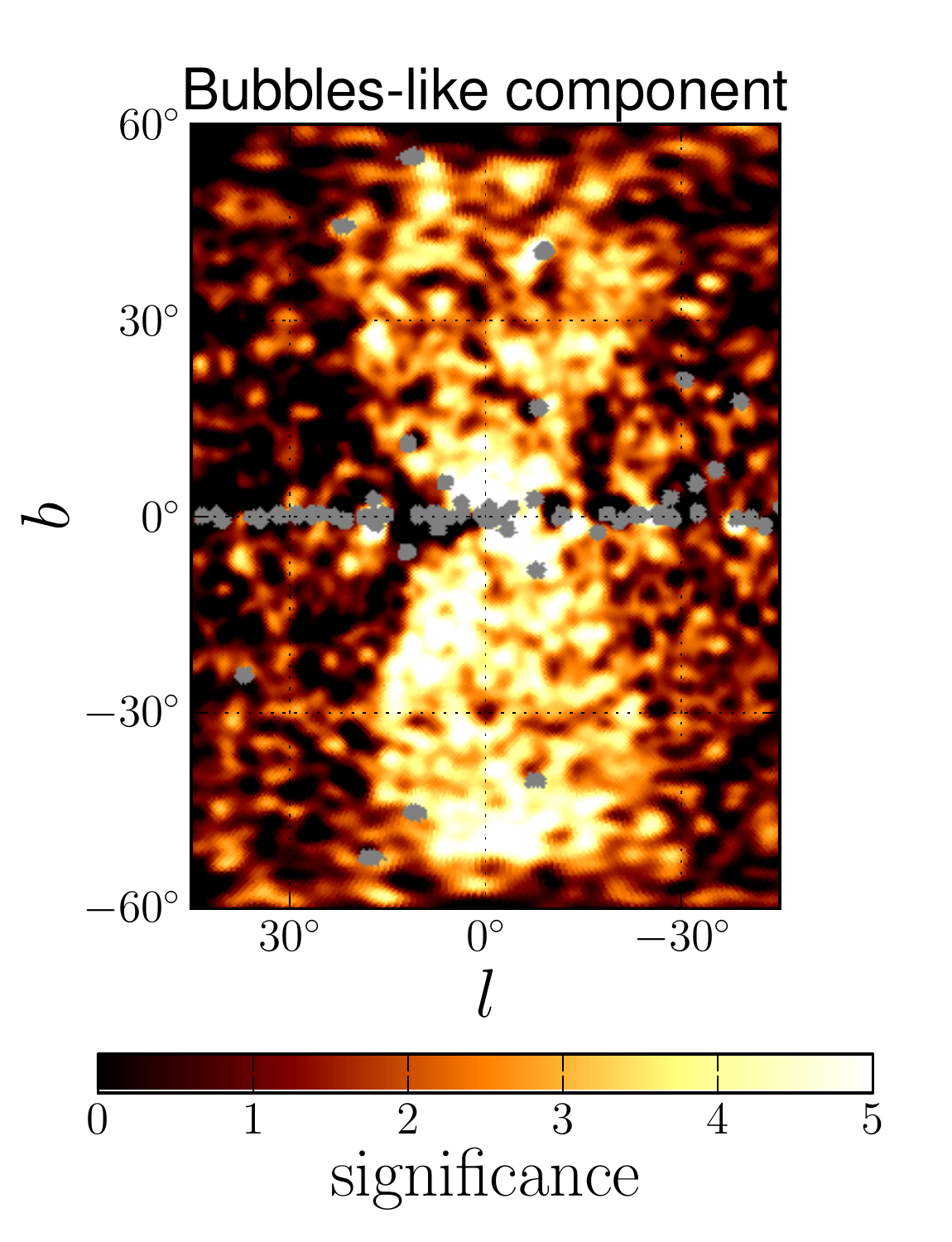}
\includegraphics[scale=\onepic]{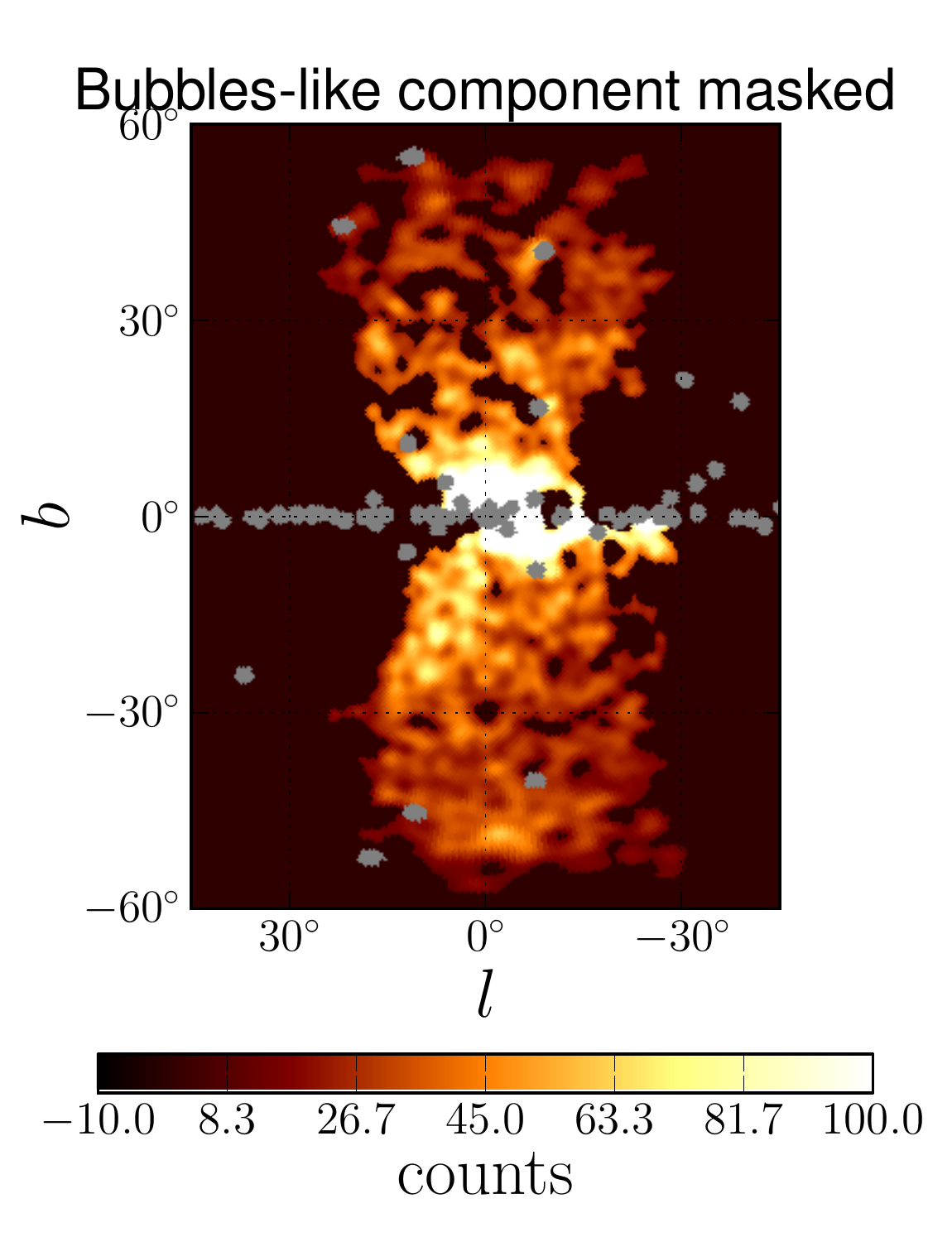}
\noindent
\caption{\small 
Left: significance map of the $E^{-1.9}$ spectral component in the middle panel of
Figure \ref{fig:SCA_components}.
Right: bubbles template obtained from the $E^{-1.9}$ spectral component
by cutting at $2\sigma$ significance in the left plot and taking the connected component.
}
\label{fig:SCA_bubbles}
\end{center}
\end{figure}

In order to derive the bubbles template, we first model the gamma-ray emission by the gas-correlated templates ($\pi^0$ and bremsstrahlung),
PS, and a combination of smooth components, which are introduced to provide a generic model for the other
components of emission, such as the Loop I and the bubbles.
Then we subtract the gas-correlated components and the point sources from the data and decompose the remaining
residuals between 1 GeV and 10 GeV into two components correlated with $\propto E^{-1.9}$ and $\propto E^{-2.4}$ spectra.
The former spectrum is the spectrum of the \Fermi bubbles at high latitudes, the latter one is the average spectrum 
of the other astrophysical components: Loop I, IC, and isotropic. The residual after subtracting the gas-correlated emission
and PS from the data and the two spectral components are shown in figure \ref{fig:SCA_components}.
Then we introduce a cut in significance at $2\sigma$ level of the $\propto E^{-1.9}$ 
component to derive the \Fermi bubbles template (Figure \ref{fig:SCA_bubbles}).

\begin{figure}[htbp]
\begin{center}
\includegraphics[scale=\onepic]{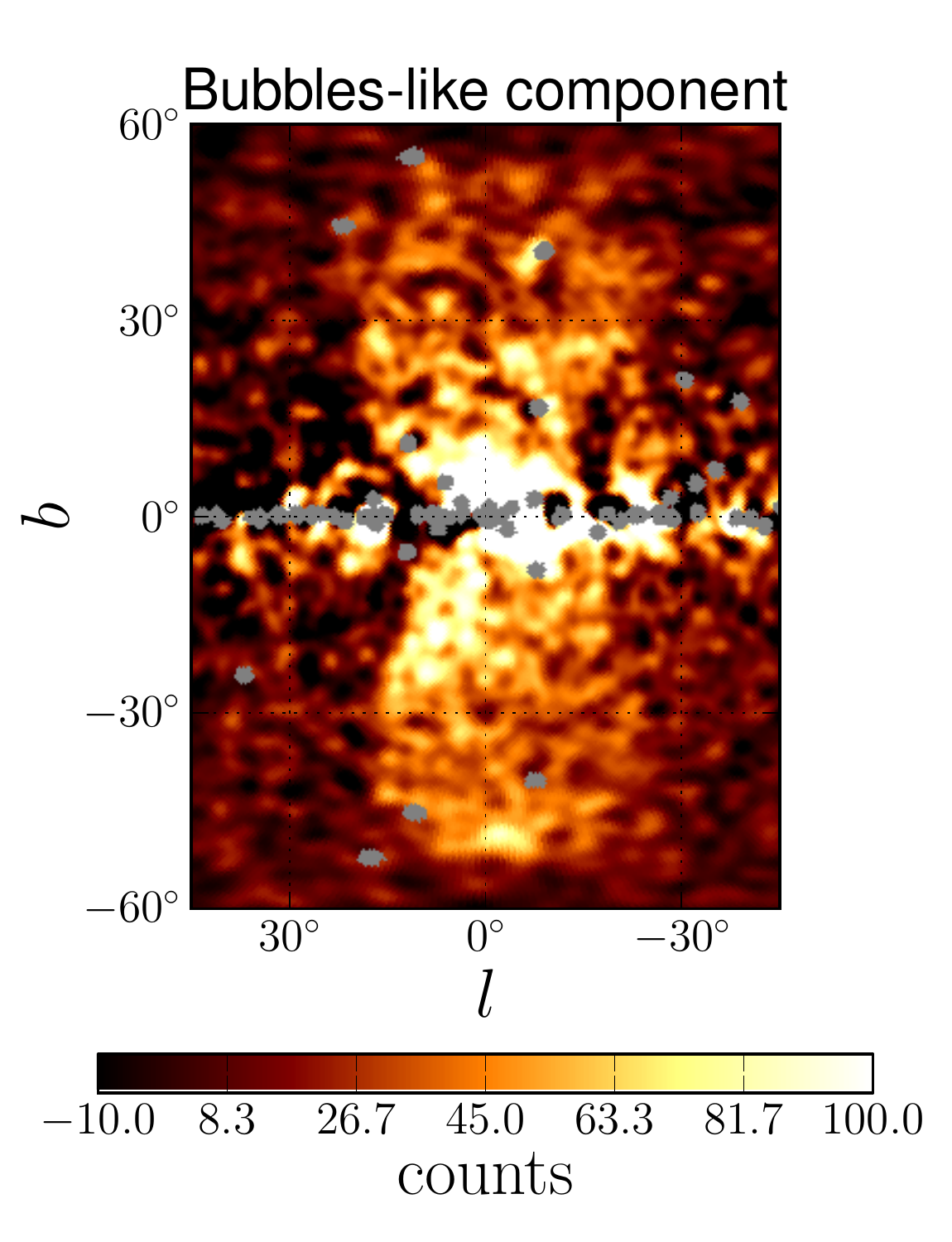}
\includegraphics[scale=\onepic]{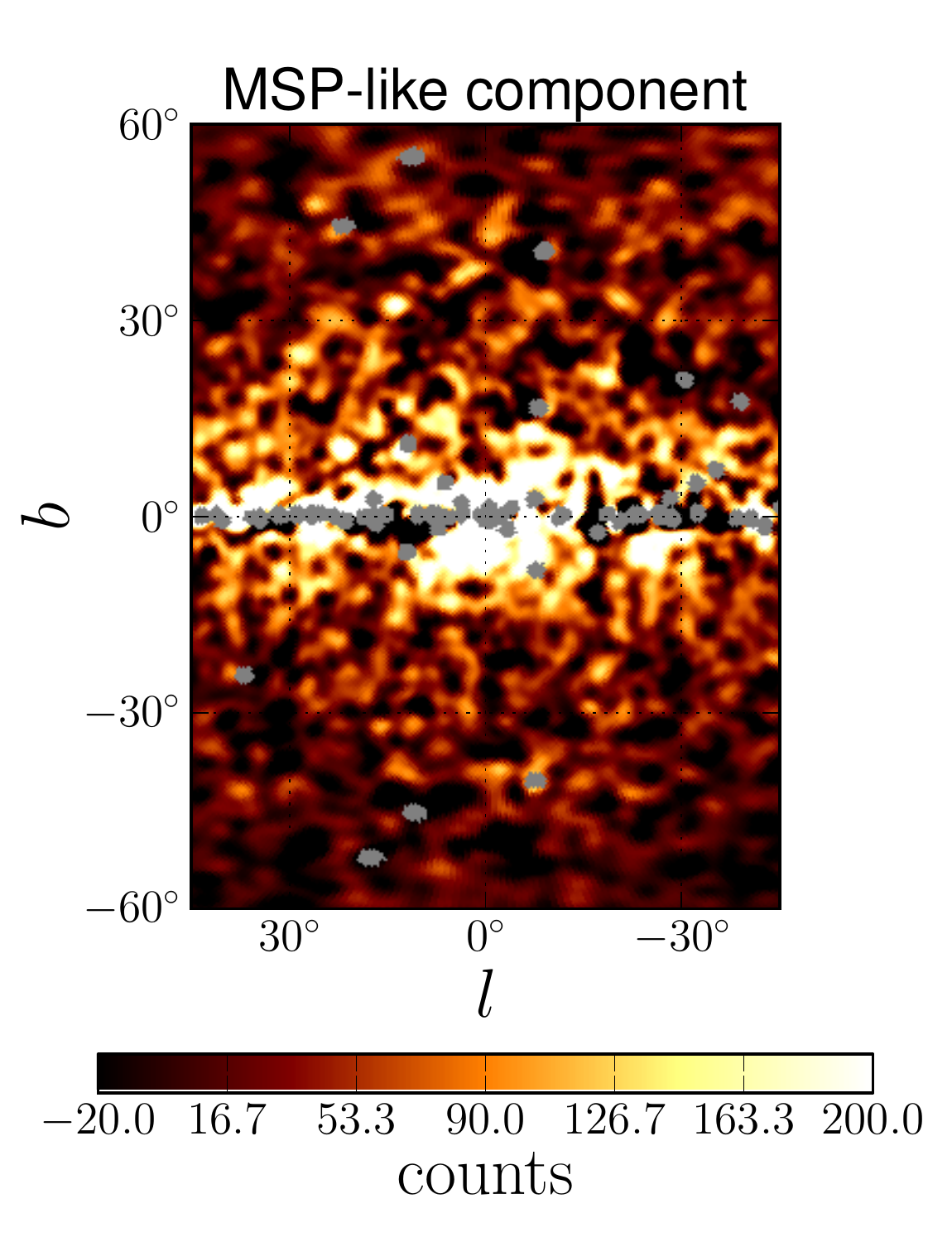}
\noindent
\caption{\small 
Bubble-like $\propto E^{-1.9}$ (left panel) and MSP-like $\propto E^{-1.6} e^{-E/4\:{\rm GeV}}$ (right panel)
spectral components obtained in three-component decomposition of the residuals 
in the left panel of Figure \ref{fig:SCA_components}.
}
\label{fig:SCA_cutoff_components}
\end{center}
\end{figure}

An alternative template of the bubbles as well as a template of the GC excess can be derived if we separate the residuals
after subtracting the gas-correlated components and PS from the data into three spectral components,
where the first two components have the same spectra as before ($\propto E^{-1.9}$ and $\propto E^{-2.4}$)
while for the third component we take an average spectrum of MSPs $\propto E^{-1.6} e^{-E/4\:{\rm GeV}}$
\cite{2014arXiv1407.5583C, 2015ApJ...804...86M}.
The maps for the bubble-like and MSP-like spectral components are shown in Figure \ref{fig:SCA_cutoff_components}.

\begin{figure}[htbp]
\begin{center}
\includegraphics[scale=\onepic]{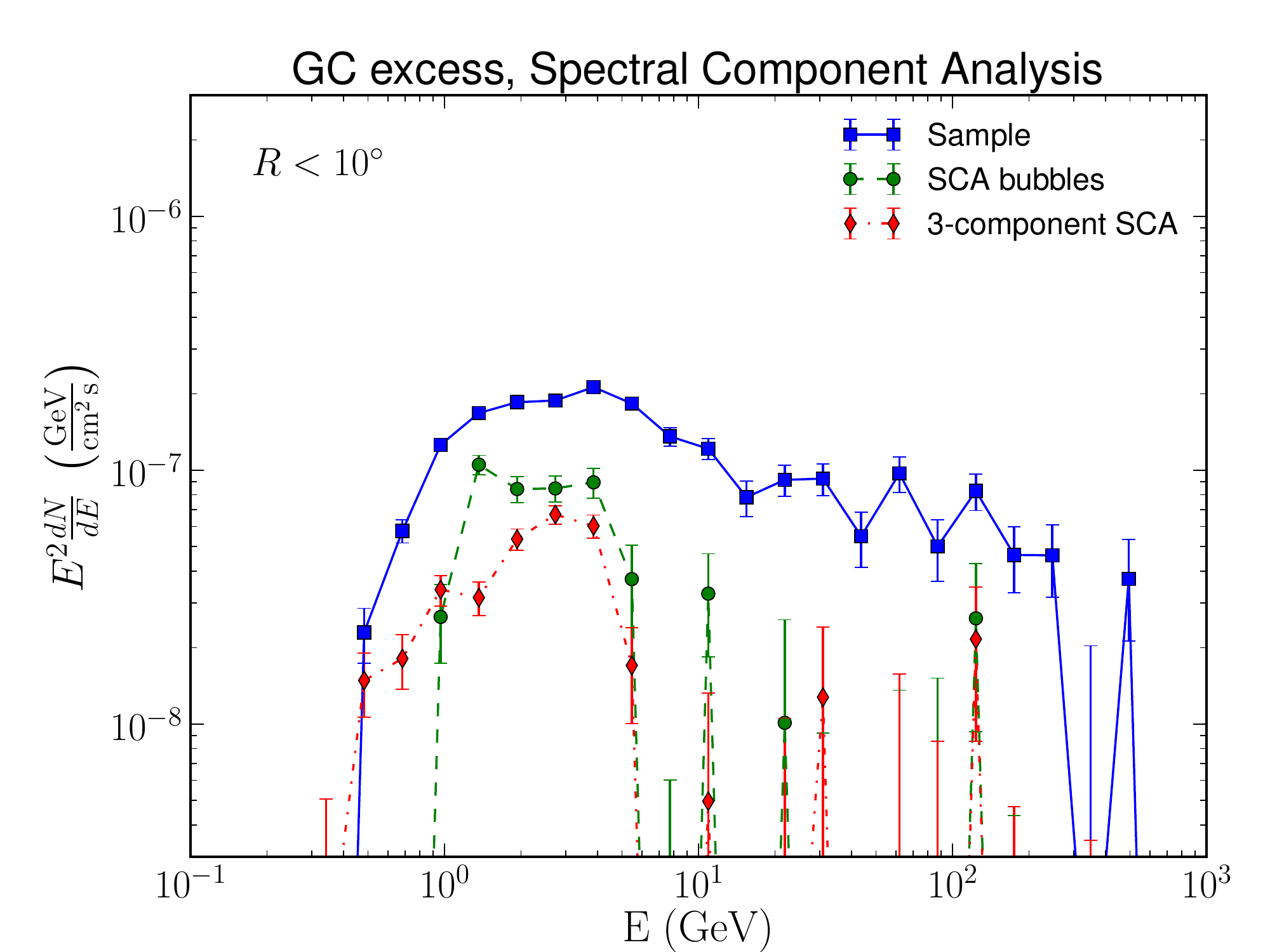}
\noindent
\caption{\small 
Spectrum of the excess integrated within $10^\circ$ from the GC. ``Sample" model corresponds to the GC excess spectrum
in Figure \ref{fig:baseline_spectra} on the left. It is derived with the \Fermi bubbles template for $|b| > 10^\circ$.
In the ``SCA bubbles", the bubbles template is presented in Figure \ref{fig:SCA_bubbles} on the right, while the GC excess is modeled
by the gNFW template. In the ``3-component SCA" model with bubbles and the GC excess templates are derived from the
``Bubbles-like" and ``MSP-like" spectral components in Figure \ref{fig:SCA_cutoff_components} respectively.
}
\label{fig:SCA_excess_spectra}
\end{center}
\end{figure}

The effect of including the all-sky bubbles template on the GC excess flux is shown in Figure \ref{fig:SCA_excess_spectra}.
We see that the GC excess flux is completely absorbed by the \Fermi bubbles template at $E > 10$ GeV
and it is reduced by a factor $\gtrsim$ 2 below 10 GeV.

\section{GC excess and \Fermi bubbles near the GC}

\begin{figure}[htbp]
\begin{center}
\includegraphics[scale=\twopic]{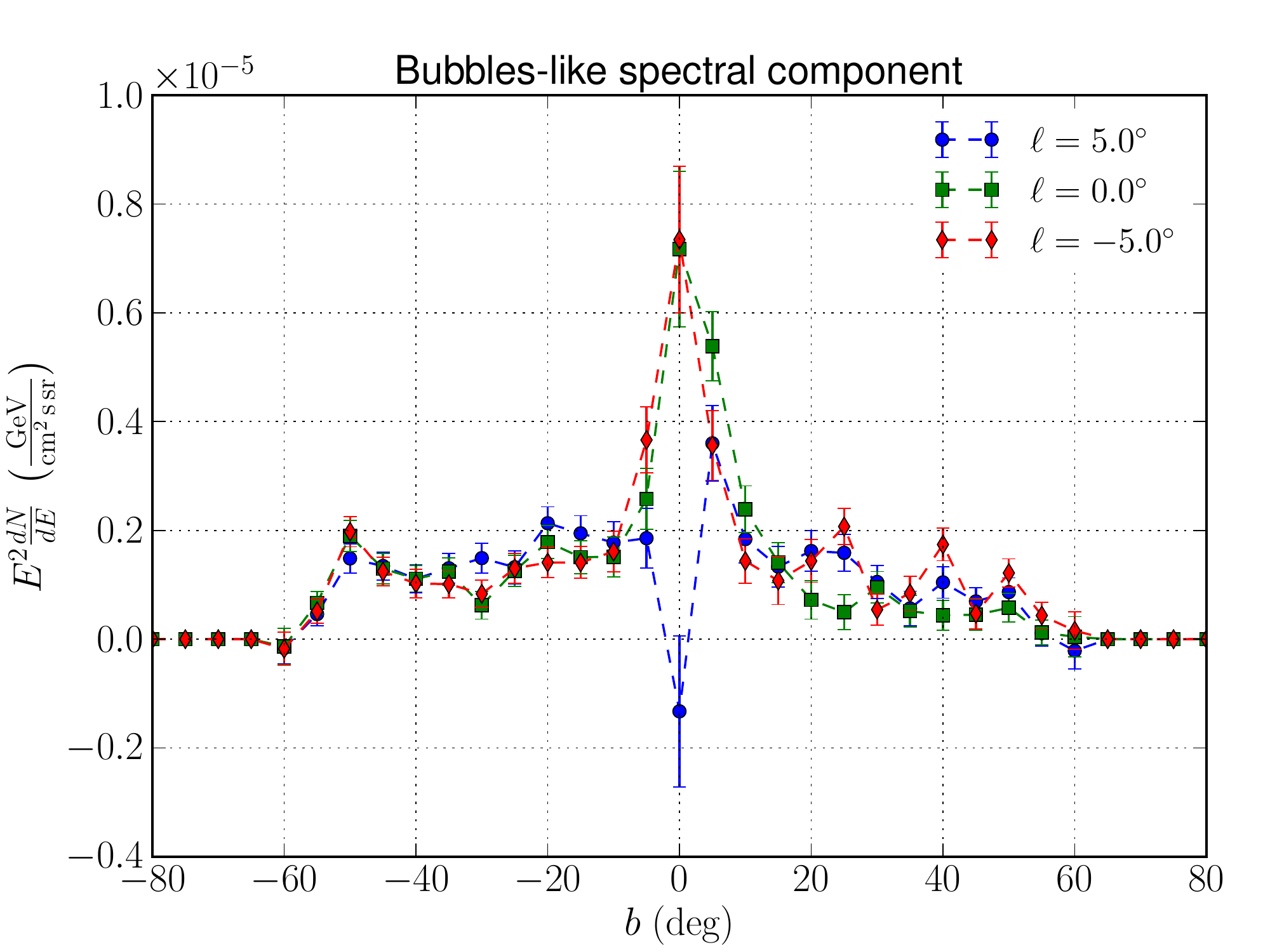}
\includegraphics[scale=\twopic]{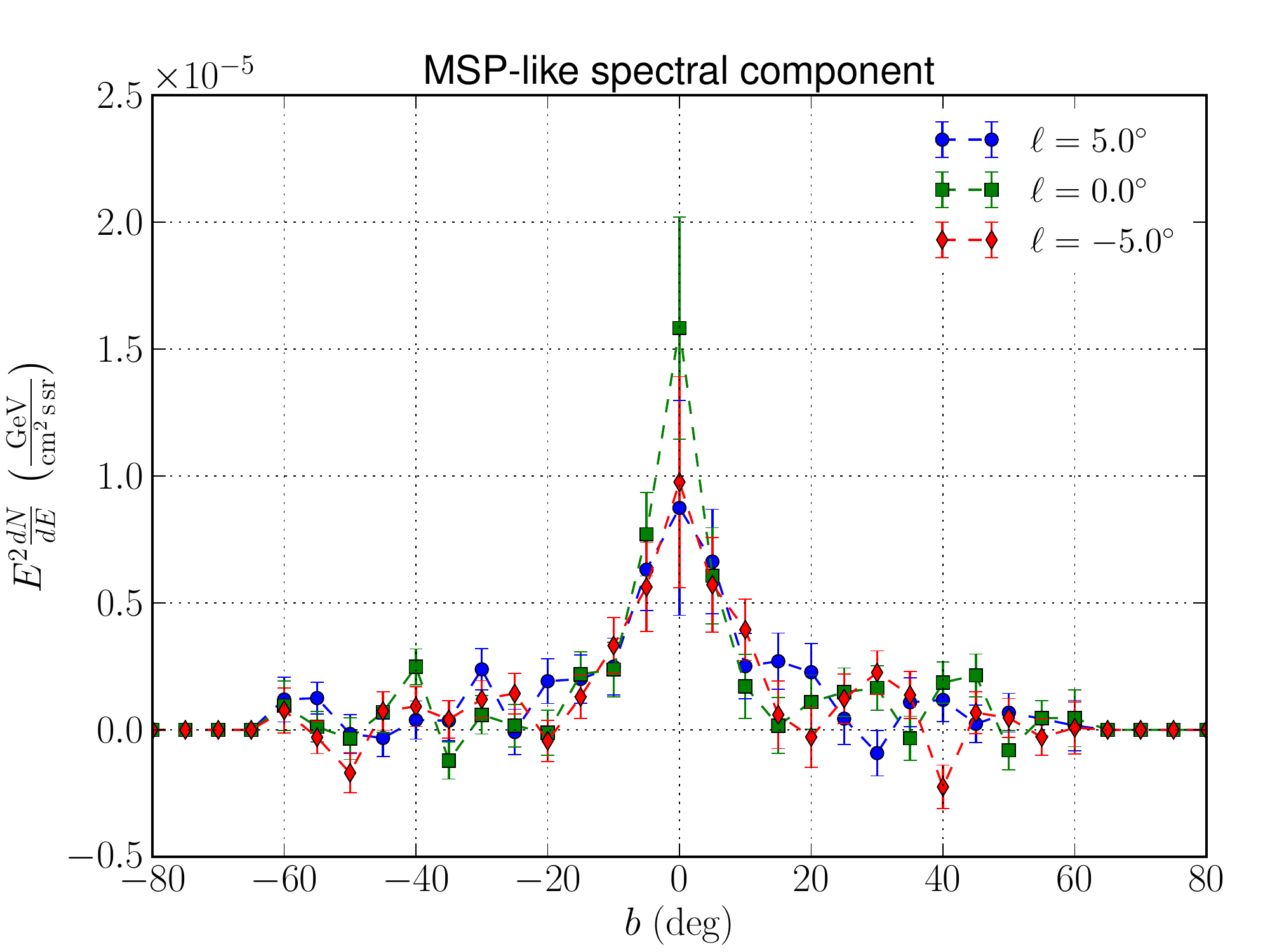}
\noindent
\caption{\small 
Latitude profiles of the bubble-like (left) and MSP-like (right) spectral components
in Figure \ref{fig:SCA_cutoff_components}.
The normalization corresponds to the intensity of these components at 2 GeV.
}
\label{fig:lat_profiles}
\end{center}
\end{figure}

In this Section we briefly discuss the morphology of the bubble-like and the MSP-like spectral components.
The latitude profile plots for the spectral components maps in Figure \ref{fig:SCA_cutoff_components} are
shown in Figure \ref{fig:lat_profiles}.

The bubble-like component is shown on the left. The distribution is flat at latitudes $10^\circ < |b| < 50^\circ$.
The flatness of the profile is especially clear at negative latitudes where there is less overlap with the local 
gas clouds compared to positive latitudes.
However, close to the GP, there is an increase in intensity of emission for longitudes $\ell = 0^\circ,\: -5^\circ$ by a factor 3 to 4 
compared to high latitudes, while the intensity around $\ell = 5^\circ$ is consistent with zero.
Thus, the intensity of emission from the \Fermi bubbles appears to have larger intensity near the GP for longitudes $\ell \lesssim 0^\circ$.
The apparent asymmetry of the \Fermi bubbles with respect to the GC may have implications for the 
interpretation of the \Fermi bubbles as emanating from the supermassive black hole at the GC.
The presence of the asymmetry is subject to large uncertainties in the distribution of gas towards the GC
\cite{2017arXiv170403910T} and needs to be investigated further.

The latitude profiles for the MSP-like spectral component exhibit a morphology consistent with spherical symmetry 
with respect to the GC.
There is a slight excess of emission in the GP at $\ell = \pm 5^\circ$, $b = 0^\circ$ 
relative to the off plane locations $b = \pm 5^\circ$, $\ell = 0^\circ$.
If this component is indeed coming from MSPs, then 
the slightly higher intensity along the GP can be interpreted as contribution of MSPs in the disk of the Galaxy
on top of the MSPs in the bulge.
One can also see the slightly larger emission along the GP in Figure \ref{fig:SCA_cutoff_components} on the right.
Overall, we find that the spectral component derived with MSP-like spectrum is generally consistent with the expectations
for the distribution of MSPs in the Galaxy.

\begin{figure}[htbp]
\begin{center}
\includegraphics[scale=\onepic]{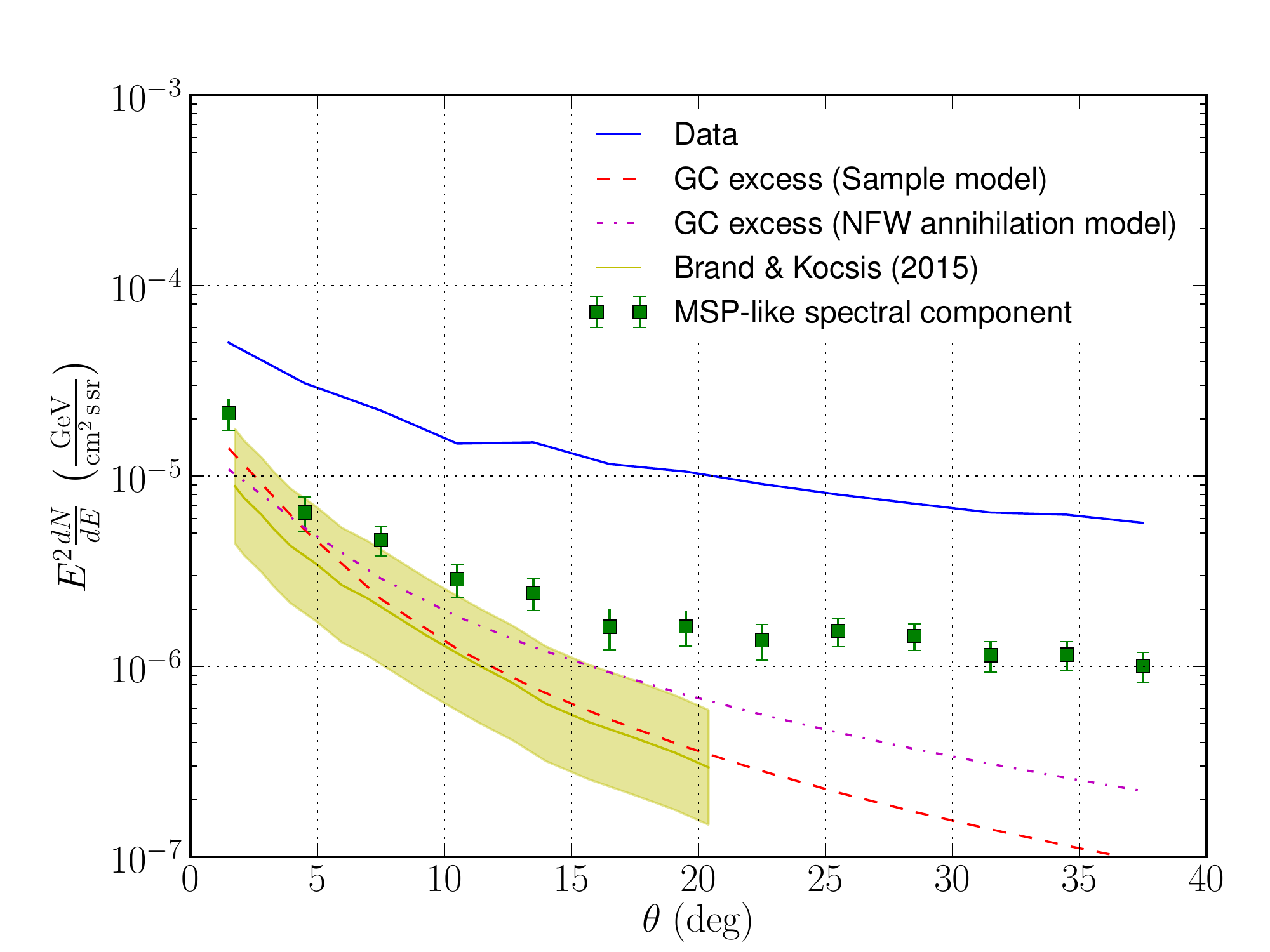}
\noindent
\caption{\small 
Radial profile of the MSP-like spectral component in Figure \ref{fig:SCA_cutoff_components} on the right.
}
\label{fig:MSPlike_radial}
\end{center}
\end{figure}

The radial profile of the MSP-like component is shown in Figure \ref{fig:MSPlike_radial}.
Within $\approx 7^\circ$ the profile is consistent with the GC excess modeled by the gNFW profile (Section \ref{sec:sample}).
The deviation at large distances from the GC can be interpreted as the contribution of the MSPs in the GP.

\section{Conclusions}

In this report we discuss a derivation of the \Fermi bubbles template at low latitudes and 
the effect of the inclusion of this template in the model for Galactic emission on the spectrum of the GC excess.
We find that the \Fermi bubbles near the GP may have an intensity 3 to 4 times higher than at high latitudes.
The bubbles in the GP may also have asymmetric distribution with respect to the GC:
in our model, the intensity of the \Fermi bubbles at positive latitudes is significantly smaller than the intensity at negative
latitudes.
We also find that modeling of the \Fermi bubbles near the GP can have a significant impact on the properties of the GC
excess. 
In the presence of the all-sky bubbles template, 
the GC excess flux is consistent with zero above 10 GeV and is reduced by a factor of about 2 or more below 10 GeV
compared to the model with the bubbles template determined for $|b| > 10^\circ$.

Future observations, such as the Cherenkov Telescope Array may be able to detect the higher intensity
emission from the \Fermi bubbles in the GP.
If the spectrum of the \Fermi bubbles in the GP does not have a softening or a cutoff observed at high latitudes
\cite{2014ApJ...793...64A},
then it may be possible to detect neutrino emission from the \Fermi bubbles 
with IceCube and KM3NeT
in the hadronic model of gamma-ray emission.
Further study of the microwave counterpart of the \Fermi bubbles at low latitudes
may help to discriminate the leptonic and hadronic models of the gamma-ray emission.
Progress in understanding of the \Fermi bubbles near the GC will help to 
get a more precise description of the GC excess, which will help to disentangle its nature.

\acknowledgments
The \textit{Fermi}-LAT Collaboration acknowledges support
from NASA and DOE (United States), CEA/Irfu, IN2P3/CNRS, and CNES (France), ASI, INFN, and INAF (Italy), MEXT, KEK, and JAXA (Japan), and the K.A.~Wallenberg Foundation, the Swedish Research Council, and the National Space Board (Sweden). 
This work was partially supported by NASA grants NNX14AQ37G and NNH13ZDA001N.